# ATM Networking: Issues and Challenges Ahead

Raj Jain
Department of Computer and Information Science
The Ohio State University
2015 Neil Avenue
Columbus, OH 43210-1277
Email: Jain@ACM.Org

Abstract

The paper begins with a discussion of current trends in networking and a historical reviews of past networking technologies some of which failed. This leads us to the discussion about what it takes for a new technology to succeed and what challanges we face in making the current dream of a seamless world-wide high-speed ATM network a reality.

Issues in using ATM cells for very high speed applications are presented. Ensuring that the users benefit from ATM networks involves several other related disciplines. These are reviewed.

## 1  Trend: Networking is Critical

Networking has become the most critical part of computing. Today, computers are used mostly for transferring information from one peripheral to another, from network to the disk, from disk to the video screen, from keyboard to the disk, and so on. Mail, file transfer, information browsing using World Wide Web, Gopher, and WAIS takes up more time of the computing resources than computing per se. Initially, when the computers were designed, the performance was measured by the "add" instruction time. Today, it is the "move" instruction that is the key to the perceived performance of a system. This means that the bus performance is more important than the arithmetic logical unit (ALU) performance. I/O performance is more important than the SPECmarks.

There are several other reasons for communications and networking becoming critical. First, the users have been moving away from the computer. In the sixties, computer users went to computer rooms to use them. In the seventies, they moved to terminal rooms away from the the computer rooms. In the eighties, the users moved to their desktop. In the nineties, they are mobile and can be anywhere. This distance between the users and the computers has lead to a natural need for communication between the user and the computer or the user interface device (which may be a portable computer) and the servers.

Second, the system extent has been growing continuously. Up until eighties, the computers consisted of one node spread within 10 meters. In nineties, the systems consists of hundreds of nodes within a campus. The increasing extent leads to increasing needs for computing.

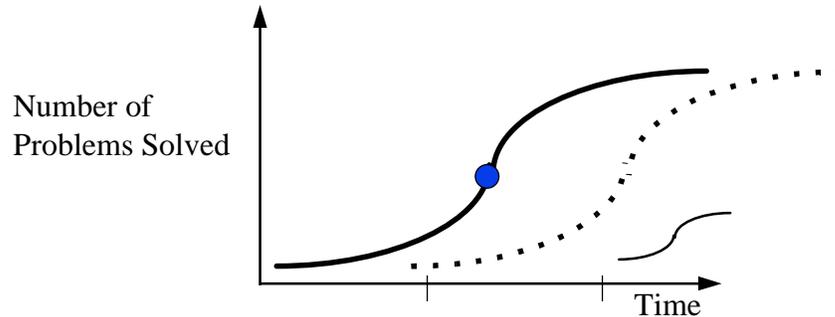

Figure 1: S-Shaped technology curve

In the last ten years, we have seen increasing personalization of computing resources. We moved from timesharing to personal computing. Now we need ways to work together with other users. So, in the next ten years, emphasis will be on cooperative computing. This will further lead to increase in communications.

In the last decade, we were busy developing corporate networks, and campus networks. In the next decade, we need to develop intercorporate networks, national information infrastructures, and international information infrastructures. All these developments will lead to more growth in the field of networking and more demands for the personnel with skills in networking.

The increasing role of communications in computing has lead to the merger of the telecommunications and computing industries. The line between voice and data communications is fading away. Data communication is expected to take over voice communication in terms of volume as discussed in Section 6.4.

## 2   Trend: Peak of Technology Life Cycle

Most technologies follow an S-shaped curve shown in Figure 1, where the number of problems solved is plotted against time. There are three distinct stages in the life of a technology. In the beginning, all problems are hard and it takes a lot of resources and time to solve a few problems. At this stage, a lot of money is spent in research but there is very little revenue. Most of this research is funded by the traditional government funding agencies, such as, National Science Foundation and Advanced Research Project Agency (ARPA) in the United States.

After some of the key problems have been solved, a lot of other problems can be solved by spending little money. At this stage, the curve takes an upturn. The amount of revenues to be made from the technology is much more than the investments. It is at this stage, that the industries take over technology development. Numerous small companies are formed and quickly grow to become large corporations.

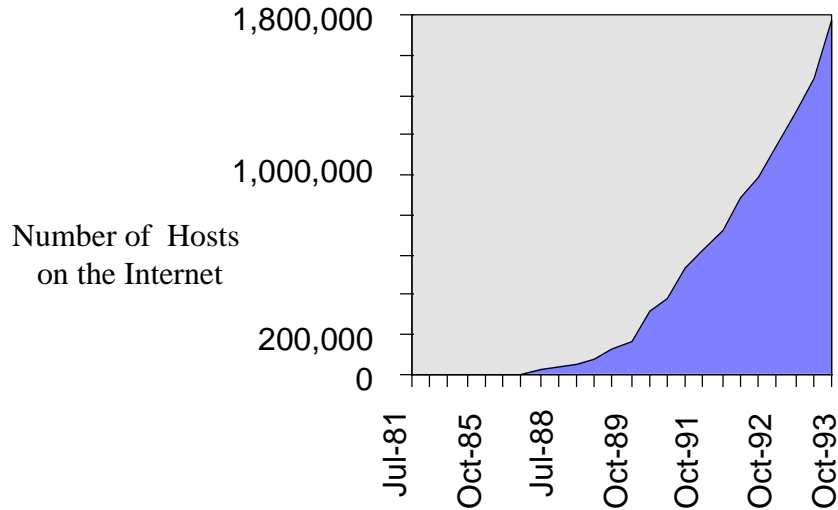

Figure 2: The exponential growth of Internet

Finally, when all the easy problems have been solved, the remaining problems are hard and would require a lot of resources. At this stage, the researchers usually move on to some other technology and a new S-shaped curve is born.

The computing industry in general and the networking sector in particular is currently going through the middle fast growing region of the technology life cycle curve. The number of problems solved is indicated by the deployment of the technology. In case of networks, one can plot the number of hosts on the networks, bytes per host, number of networks on the internet, total capacity (in MIPS) of hosts on the network, total memory or total disk space of the hosts on the network. In each case, one would see a sudden exponential up turn in the last few years.

Figure 2 shows the famous Internet growth curve. The figure shows the number of hosts on the Internet in the last 20 years. The data before July 1988, although plotted is hardly visible. Since 1988, growth has defied all predictions.

# 3   Trend: Standardization

When a technology reaches the middle fast growing region, it becomes necessary to standardize it to make it usable for the masses. The computing industry in general and networking in particular is undergoing through this phase. Even if people use different computers, it is necessary that the networking interfaces be standardized so that these diverse computers can communicate with one another.

The standardization requires a change in the way business is done. Before standardization, a majority of the market is vertical. The only way for users to maintain compatibility is to buy the complete system from one manufacturer. System vendors make more money than component vendors. IBM, DEC, and Sun Microsystems are examples of such system vendors. After standardization, the business situation changes. Users can and do buy components from different vendors. The market becomes horizontal. Companies specializing in specific components and fields take prominence. Intel for processors, Microsoft for operating systems, Novell for networking are examples of this trend.

To survive in this post-standardization era, invention alone is not sufficient. Only those new ideas that are backed by a number of vendors become standardized and are adopted. It, thus, becomes necessary to form technology partnerships.

# 4 Past Failures and Successes

In the last fifteen years, we have seen a number of networking technologies that were very promising during their life time but were not successful in the long run. A sample of such technologies is listed in Table 1. In each case, the technology listed in the second column was more promising than the one in the third column until the year shown in the first column.

In early 80's, when Ethernet was being introduced, some argued that broadband Ethernet, which allows voice, video, and data to share a single cable would be more popular than baseband Ethernet. As we all know, today there are a few broadband installations. Most installations of Ethernet are baseband. The cost of combining the three services was just too high. The analog circuits required for frequency multiplexing were not as reliable and economical as digital circuits with separate wiring.

Around the same time, when computer companies were trying to sell Ethernets, PBX manufacturers were presenting PBX as the better alternative, again because it was already there and it could handle voice as well as data. However, PBX was not accepted by the customers simply because it did not provide enough bandwidth.

The Integrated Service Digital Network (ISDN) was standardized in 1984 and was very promising then. However it's deployment has been much too slow. Even after ten years, it is not possible to get an ISDN connection at most places. Even at those places where it is available, the 64 kbps bandwidth provided by it is not sufficient for most data applications. For low bandwidth applications, modems on analog lines provide a better alternative. Modem technology has advanced much beyond expectation. Today, one can get 28.8 kbps and 56 kbps modems that work with all pervasive analog lines and do not have monthly charges associated with the extra ISDN line.

In 1986, IEEE 802.4 (token bus) was touted as a better alternative than IEEE 802.3/Ethernet for real time environment. It was said that Ethernet could not provide the delay guarantees required for manufacturing and industrial environments. Manufacturing Automation Protocol (MAP/TOP) was seen as the right solution. Today, IEEE 802.3/Ethernet is used in all such environments. Token buses are practically nonexistant.

Table 1: Networking Failures and Successes of the Past

| Year | Failure | Success |
|------|---------|---------|
| 1980 | Broadband Ethernet | Baseband Ethernet |
| 1981 | PBX | Ethernet |
| 1984 | ISDN | Modems |
| 1988 | OSI | TCP/IP |
| 1991 | DQDB |  |
| 1992 | XTP | TCP |

Up until 1988, ISO/OSI protocol stack was seen as the leading contender for networking everywhere. Networking researchers in most countries were implementing ISO/OSI protocols. The United States Government Open Systems Interconnection Profile (GOSIP) even made ISO/OSI a mandatory requirement for government purchases. Today, TCP/IP protocol stack dominates instead. The OSI protocols suffered from the common problems of standards: it had too many features. Any feature required by any application in the world needed to be supported by the standard. The protocols took too long to standardize and were quite complex. The "build before you standardize" philosophy of the TCP/IP protocol stack helped in its success.

Up until 1991, IEEE 802.6 Dual Queue Dual Bus (DQDB) was seen as a promising candidate for metropolitan area networks. It is no longer considered viable. The unfairness problem and general problems of bus architectures have made it undesirable.

Xpress Transfer Protocol (XTP) was designed as the high-performance alternative to TCP/IP. Protocol Engines – the company leading the design of XTP declared bankruptcy in 1992.

## 4.1 Requirements for Success

There are several lessons to learn from the list presented in Table 1. First, all technologies appear very promising when first proposed. However, not all survive. Those that survive meet all the following requirements.

1. Low cost
2. High performance
3. Killer application

4. Timely completion

5. Interoperability among various implementations of the same technology

6. Coexistence with existing (legacy) technology.

After a brief overview of ATM networks in the next section, we discuss some of these issues as they apply to ATM networks.

# 5  ATM Networks: An Overview

In this section we summarize the key features of ATM networks and their implications:

1. **Asynchronous**: ATM stands for Asynchronous Transfer Mode. The key innovation here is the word Asynchronous – non-periodic. The telephone networks today use synchronous transfer mode. The clocks at successive nodes (switches) of the networks are synchronized and the information transfer is periodic. Users are assigned slots and every $n$th slot belongs to the same user. This mode, also known as time-division multiplexing, matches well with circuit switching applications where information arrives periodically. However, for data applications, it is better to be able to transmit asynchronously. This is similar to packet switching and is more suitable for high-speed.

2. **Short Fixed Size**: ATM cells are all fixed 53 bytes long. As discussed in Section 6.5, this reduces the variance in delay and helps in delay critical applications.

3. **Connection-Oriented**: All switches on the path are consulted before a new connection is set up in an ATM networks. The switches can ensure that they have sufficient resources before accepting a connection. It is much easier to guarantee quality of service in a connection-oriented network than in connectionless networks. The connections in ATM networks are called Virtual Circuits (VCs).

4. **Labels vs Addresses**: ATM networks use VC Ids (VCIs) to identify cells of various connections. VCIs are assigned locally by the switch. A connection has different VCI at each switch. The same VCI can be used for different connections at different switches. Thus, the size of VCI field does not limit the number of VCs that can be set up. This makes ATM networks scalable in terms of number of connections. In TCP/IP networks, the number of end systems is limited by the size of the address fields making them non-scalable in terms of number of connections.

5. **Switching vs Routing**: Switching fixed size cells based on short VCIs is much simpler than routing of variable size packets based on addresses. This makes ATM networks better scalable in terms of bandwidth.

6. **Seamless**: In addition to the number of connections and speed, ATM technology is also scalable in terms of distance. The same technology can be used for LANs or WANs.

In summary, ATM technology has better scalability and lower delay variance than current packet switching technology.

# 6 Challenges

In this section, we discuss a few of the requirements identified earlier in Section 4.1 and see what needs to be done to ensure success of the ATM technology.

## 6.1 Economy of Scale

Today, networking technology seems to be far ahead of the applications. High-speed fibers have been installed but there is not enough video traffic to fill them. Invention is becoming the mother of necessity. We need to create a need for the high bandwidth. In such a situation, generally low cost is the primary motivator. When the high-speed technology is proportionately cheaper than lower-speed alternatives, the buyer's considerations change from "Why should I buy high speed?" to "Why should I not buy high speed?"

Today, there is diseconomy of scale. Higher speed networks cost more in per-bit than lower speed networks. Ten Mbps Ethernet cards can be had for $50. However, 100 Mbps FDDI cards cost closer to $1000. This diseconomy of scale has a significant impact on user adoption. We have seen this happen in other areas of computing. Today ten 100-MIPS computers cost much less than one 1000-MIPS computer. Therefore, we see more distributed computing than supercomputing. Applying the same logic, it appears that unless there is economy of scale, people may divide their networking applications among multiple low-speed links than one high-speed link. Of course, there are a few applications that will not work on speeds in the range of 10 Mbps. For these, the users have no choice but to use higher speed links.

This diseconomy of scale affects all high speed technologies including ATM. However, ATM has a bigger uphill battle due to its newness. In a recent ATM Forum user survey conducted Dr. John McQuillan, the users were asked that given the same price, which 100 Mbps network they would buy: ATM or 100-Mbps Ethernet. The answer was Ethernet because it is something the users feel very comfortable with. We ATM designers will have to work hard to get the ATM equipment prices below the 100-Mbps Ethernet to get acceptance.

## 6.2 Tariff

Tariffing the ATM traffic is another problem. Today's telecommunications tariffs are designed for low bandwidth high cost voice traffic. It costs $25/month for a simple 64-kbps

analog phone line. At this rate, the phone company has the potential of making $211k/year on a 45 Mbps link. A coast to coast T3 link does cost $180k to $240k per year. A 155 Mbps link would cost three times as much. However, 155 Mbps ATM circuits are being tariffed at $13k to $45k per year. That's 10 to 50 times cheaper than today's rates. The situation is similar to that of the computer industry. When the computer prices started going down, the old established companies designed to sell expensive mini and supercomputers had trouble keeping their overheads down to be able to sell personal computers. While new companies designed to make money selling personal computers cheaply flourished, old companies tried hard to maintain their existing business almost to the point of bankruptcy. Success of cheap telecommunications using ATM technology has the potential of doing the same to the telecommunications industry. The telecommunications companies with overheads designed for expensive voice services will have a tough time selling cheap data services. The danger of decreasing revenues may prove to be a hinderence to the success of ATM at least in the wide area networking (WAN) market.

## 6.3 Performance

Figure 3 shows a layered model of people involved in designing a high speed network. At the bottom are the physicists, who work on fibers and lasers. They are working today with 10 Gbps lasers and their challenge is to design 100 Gbps lasers. The next level up are the LAN designers who are designing FDDI and ATM networks using these fibers. The LAN designers are working at 100 to 155 Mbps – two orders of magnitude lower than the physicists. Then there are LAN adapter designers who take the 100 Mbps protocol standard and design adapters which run barely at 20 Mbps. Although some users aren't aware of this, many FDDI adapters cannot transmit or receive more than 20 Mbps. Thus, we loose the performance by a factor of 5 at this layer.

The next (fourth) layer consists of processor designers. Unless designed carefully, a processor may not be able to keep up with the high speed LAN adapters. The fifth layer of operating system reduces the performance further. The sixth layer is that of network protocols– some of which are not able to cope with high speed links. Finally the top (seventh) layer is the application, which sees a usable bandwidth of a few megabits per second.

Unless higher layers are improved, changing the lower layers will not result in any higher performance for the user. Today's problem is not so much in networking protocols as it is in proper I/O designs for the processors and operating systems. We can provide the user with 155 Mbps or 622 Mbps links, but they will not be able to use it unless, operating system and processor designs are improved accordingly. The only exception is the backbone where specialized hardware and software are used. The backbone components (switches, routers, or bridges) are designed specially for high communication speeds. Thus, the first place where high-speed is will be used is in the backbone. The desktop market will have to wait for better operating systems and processors.

Another lesson to learn from this layered model is that for high speed networks to become a reality, all seven layers have to be improved. Bad performance even in one layer can delay

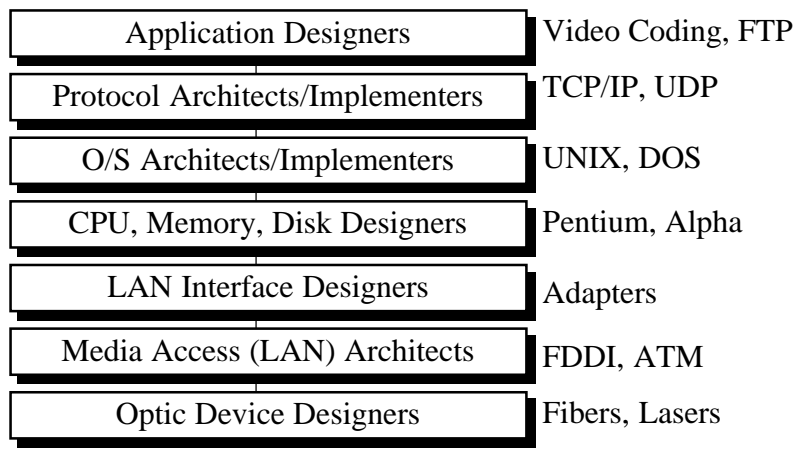

Figure 3: A layered view of people who affect the performance seen by the user

the introduction of the high-speed networking.

## 6.4 Application

It is well known that future applications will be multimedia including data, voice, image, and video. What is less well known is that the voice traffic will be a negligible part of the traffic. This can be seen by considering what happens on telephone networks today. On the AT&T network, approximately 125 to 130 million calls are made per day and an average call lasts around 5 minutes. Each call requires a bandwidth of 64 kbps. Thus, the total bandwidth used by voice in AT&T network is approximately 28.8 Gbps, which is only one thousandth of the potential bandwidth of a fiber. Even if all 200 million people of the United States were to talk 24 hours a day, the total bandwidth required will be only 12.8 Tbps, which again is less than the potential bandwidth of a fiber. In a survey of private networks reported in the August 1992 issue of IEEE Spectrum, it was found that in 1985, 75% of the trafic was voice. In 1990, the voice percentage dropped to 56% and in 1995, it is projected to be 39%. If we were to make a forecast based on this trend, we would conclude that the voice traffic will be zero or negligible by the year 2010. The reason for this decline is not that people are not talking enough but that while voice traffic is limited by the population, the data traffic is not. Computers have no limit on the speed at which they can transmit and so there is no limit to the value to which data traffic can grow. It is this exponentially growing data business that most telecommunications companies want to get into.

Next, let us consider characteristics of video traffic. One hour of uncompressed HDTV requires 540 GB of storage. At today's storage prices of \$1/MB, this works out to approx-imately $150 per second of video. This is somewhat expensive and only researchers funded on government grant can afford to store the video at this price. If compressed, the storage requirements drop by a factor of 60 to 200 and the price becomes $2.50 to $0.75 per second, which is more reasonable. The conclusion is that most video will be in compressed form simply for storage. Compression means that the bandwidth requirements vary, and therefore, variable bit rate (VBR) service rather than constant bit rate (CBR) is likely to be used more often.

Also, at high speeds, the connection holding times become shorter. At 1 Gbps, it takes only 10 seconds to transmit 1 hour of compressed VHS movie. It takes even smaller time at higher speeds. Thus, unless the bandwidth is free, most users of high speed will start and shutdown the connection after 10-20 seconds. In other words, the traffic will be short-lived and bursty. This is closer to today's data traffic than voice traffic. For ATM networks to succeed, they should be able to handle the bursty traffic efficiently.

In 1984, when the ATM cell size was being decided in CCITT, they were thinking about 64 kbps voice. At that speed, 32-byte cells need 4 ms. If larger cells were used, the time to collect the voice would become too large and would require echo cancellation. The Europeans, therefore, wanted 32-byte cells, while US position was that the cells be at least 64 bytes long. The limit of 48 bytes was chosen as an average of 32 and 64. In other words, the cell size was chosen not for high speed applications but for 64-kbps voice applications. Several other design and implementation decisions for ATM networks were similarly done as if they were being designed for voice. One example of such design philosophy is simply dropping cells on congestion. Requiring users to indicate which cells aren't important by the congestion-loss priority (CLP) bit is another example. For voice, some cells can be dropped without significant impact. However, this is not true for data. Every single bit is important and all dropped bits have to be retransmitted.

The cell size is not suitable for high-speed applications in general and for video traffic in particular. A single HDTV frame requires 50,000 cells. Switching 50,000 times for each frame is a not the optimal way.

Prior to the formation of the ATM Forum in October 1991, most of the ATM networks design decisions were made as if the network was being designed for voice. It is only in 1994 that the importance of data traffic was realized and the available bit rate (ABR) service was introduced. It is now well understood that the key to ATM technology success is its support of data traffic. If ATM fails to support data, it will not be able to stay around for video traffic.

## 6.5 Scalability

Queueing theory tells us that the variance of response time in a queue depends upon the variability of the service time (cell time) and square of the cell time:

$$\text{Variance(response time)} = \text{Variance(cell time)} + \text{Cell time}^2$$

Making the cells same size makes the first term zero. Making the cell time small reduces the variance in response time. For delay sensitive real time applications, smaller cells provide reduced delay variation.

It is important to note that we used time in the above equation and not size. The time requirements of an application do not change as the bandwidth of the network changes. For example, 30 frames per second video will need one frame every 33 ms regardless of the speed of the link. Even at Gigabit per second or Terabit per second speeds, the video will need a response time variation in milliseconds. The cell time of 6 ms would satisfy most delay sensitive applications. Although 6 ms is 48 bytes at 64 kbps, it is 900 kB at 1.2 Gbps. By using smaller cells at higher speeds we get micro- and nano-second delay variation. Unfortunately our eyes cannot even feel the difference between millisecond and microsecond delays, and therefore, we are wasting switching resources.

The cost of a switch depends partly upon its switching speed in cells per second. Given a switch design, it is possible to make a higher speed switch by simply increasing the cell size without any significant increase in cost.

What we need are "constant-time" cells and not "constant-size" cells for scalability to high speed. With constant size cells, the cell time decreases as the speed increases and it becomes necessary to switch more and more cells per second thereby increasing the cost.

The telecommunications industry claims SONET to be scalable in bandwidth. SONET uses constant-time frames. All SONET frames are 125 microseconds long. As the speed increases, the number of bytes in the frame increases proportionately.

Many of today's ATM networks use SONET links. In these networks, the large video image is broken down in small cells, which are then packed into a large SONET frame and transmitted. At the receiver, the SONET frame is unloaded, the information is switched cell by cell – all of whom are probably going to the same destination. After switching, the cells are loaded into another SONET frame and forwarded to the next switch. This process of unloading SONET frames and switching cell by cell is clearly unnecessary, given that at high speed, the amount of information to be transmitted is also generally higher. We could just switch SONET frames or use a technology which makes use of the best features of SONET and ATM.

## 6.6 Simplicity

The final challenge that ATM technology faces is that of keeping it simple. During the design of IEEE 802.3/Ethernet standard, there was fierce competition between CSMA/CD and token ring camp - both trying their best to keep their design more cost effective than the other. This competition was good and did help keep the scope of both standards limited. For ATM, unfortunately, there is no equal competition at this time. Thus, any thing that needs to be done by networks has to be done by ATM networks. The design is becoming too complex. Too many options are being added. ATM networks have to work for constant bit rate (CBR) traffic as well as variable (VBR) and available bit rate (ABR) traffic. They have to work for local area (LAN) as well as for wide area (WAN). They have to work for low-speed as well as for high speed. They have to work for private networks as well as for public networks. All these options add to the complexity. The situation is similar to that of OSI.

One of the advantages of switches over routers was that switches were supposed to simple. They no longer are. For ATM switches, switching is only a negligible part of their responsibility. A large part of switch resources is consumed by connection set up, route determination, address translation, multicasting, anycasting, flow control, congestion control, and so on.

Another element adding to the complexity of ATM is the fact that it is being developed at multiple standards bodies: ITU and ATM Forum. Strictly speaking, ATM forum is not a standards body. ITU is supposed to develop the standards and ATM Forum is supposed to select a subset of options provided by ITU. But in reality, ITU is too slow. ATM Forum cannot wait for ITU to finalize its standard and so it is taking a leading role in developing it in parallel. A considerable amount of time at both bodies is used in reconciling the agreements made at the other body. Vendors will end up implementing both ITU and ATM Forum versions of the standards and users will have to bear the cost even though just one set would have been fine.

# 7 Summary

Networking is a critical part of computing today and is growing exponentially. Networking is in the mid fast-growing region of the technology curve.

High speed networking will succeed if and only if there is economy of scale so that using higher speed links results in cost savings. Unfortunately, this is not case right now. We face the danger of users dividing their applications and using several low-speed links.

Considerable amount of resources are being put on ATM networks. However, its success will depend upon our being able to transfer data at a lower cost and higher performance than legacy LANs. Also, we will have to control the desire to incorporate all options at once otherwise it will become too complex.